\documentclass[twocolumn]{svjour3}          
\smartqed  
\usepackage{graphicx}
\usepackage{mathptmx}      
\usepackage{amssymb}
\usepackage{dcolumn}
\usepackage[utf8]{inputenc}
\usepackage{bm}
\newcommand{\be}{\begin{equation}}
\newcommand{\ee}{\end{equation}}
\journalname{Nonlinear Dinamics}

\begin{document}

\title{ Cycles in Asymptotically Stable and Chaotic Fractional Maps}


\author{Mark Edelman}

\institute{Mark Edelman \at
              Department of Physics, Stern College at Yeshiva University,   245 Lexington Ave, New York, NY 10016, USA \\
Courant Institute of
Mathematical Sciences, New York University, 251 Mercer St., New York, NY
10012, USA\\
Department of Mathematics, BCC, CUNY, 2155 University Avenue, 
Bronx, New York 10453
              \email{edelman@cims.nyu.edu}          
}

\date{Received: date / Accepted: date}

\maketitle

\begin{abstract}
Presence of the power-law memory is a significant feature of 
many natural (biological, physical, etc.) and social systems. Continuous and discrete fractional calculus is the instrument to describe the behavior of systems with the power-law memory. Existence of chaotic solutions is an intrinsic property of nonlinear dynamics (regular and fractional). Behavior of fractional systems can be very different from the behavior of the corresponding systems with no memory. Finding periodic points is essential for understanding regular and chaotic dynamics. Fractional systems don't have periodic points except fixed points. Instead, they have asymptotically periodic points (sinks). There have been no reported results (formulae) which would allow calculations of asymptotically periodic points of nonlinear fractional systems so far. 

In this paper we derive the equations that allow calculations of
the coordinates of the asymptotically periodic sinks. 
\end{abstract}

\section{Introduction}
\label{sec:1}

It is shown in many papers and reviews that power-law memory is present in social and economic systems (see, e.g. \cite{MachadoFinance,Machado2015,TarasovEconomic}). 
Power-law in human memory was demonstrated in
\cite{Kahana,Rubin,Wixted1,Wixted2,Adaptation1,Donkin}, where it is shown that the accuracy on memory tasks decreases as a power law  
$\sim t^{-\beta}$, with $0<\beta<1$.  Similar results for human learning are shown in \cite{Anderson}.
For biological power-law adaptation see papers
\cite{Adaptation1,Adaptation3,Adaptation4,Adaptation2,Adaptation5,Adaptation6}. On the microlevel, it has been shown that processing of external stimuli by individual neurons can be described by fractional differentiation \cite{Neuron3,Neuron4,Neuron5} and 
the orders of fractional derivatives $\alpha$ found in different types of neurons fall within the interval [0,1]. 
It should be noted that fractional maps corresponding to fractional differential equations of
the order $0<\alpha<1$ are maps with the power-law decaying memory in which the power is $-\beta=\alpha-1$ and $\beta \in [0,1]$ \cite{MEp1}. 
Fractional description of biological organ tissues follows from the fact that all considered organ tissues (brain, liver, spleen, etc.) are found to be viscoelastic. Corresponding values of $\alpha$ fall within the interval [0,2] (see, e.g., references in \cite{Chaos2015}). 

It is also well-known that most biological and socioeconomic systems are nonlinear and, in many cases, can be modeled as discrete systems. Study of nonlinear discrete fractional dynamics is the subject of this paper. Study of regular discrete nonlinear dynamics also was stimulated by the biological applications. The logistic map was introduced as the basic population dynamics model and many important general results in discrete nonlinear dynamics were obtained by studying this map \cite{May}. 

Study of nonlinear dynamics begins with the definition of periodic points and bifurcation diagrams. This is not a problem in regular dynamics, but it is known that 
continuous and discrete fractional systems may not have periodic 
solutions except fixed points (see, e.g., 
\cite{PerD1,PerD2,PerC1,PerC2,PerC3,PerC4,PerC5}. Instead, they may have asymptotically periodic solutions (see papers \cite{ME3,ME2,Chaos,ME4} and reviews \cite{HBV2,HBV4}). A formula for calculation of the asymptotically period two solutions for fractional and fractional difference maps was presented in \cite{Chaos2018} without a strict proof. In this paper we present a strict proof of the validity of the formulae for calculation of any asymptotically periodic cycles (sinks) derived for fractional and fractional difference maps. 
As in regular dynamics \cite{Cvitanovic}, asymptotically periodic orbits should represent 
the skeleton of fractional chaos.   
These formulae can be used to analyze stable asymptotically periodic solutions and chaos in discrete fractional systems.  

\section{Fractional/fractional difference maps}
\label{sec:2}

In this section we will omit an introduction of the basic definitions of fractional and fractional difference calculus and refer the reader to two relevant reviews \cite{HBV2,HBV4}. We will present only already widely accepted definitions of the Caputo fractional and fractional difference universal $\alpha$-families of maps. 

The Caputo universal map of the order $\alpha$ is defined (derived) as
{\setlength\arraycolsep{0.5pt}
\begin{eqnarray}
&&x_{n+1}= \sum^{m-1}_{k=0}\frac{b_k}{k!}h^k(n+1)^{k} 
\nonumber \\  
&&-\frac{h^{\alpha}}{\Gamma(\alpha)}\sum^{n}_{k=0} G_K(x_k) (n-k+1)^{\alpha-1},
\label{FrCMapx}
\end{eqnarray} 
}
where $\alpha \in \mathbb{R}$, $\alpha \ge 0$, $m=\lceil \alpha \rceil$,  $x_n=x(t=nh)$, $t$ is time, $n \in \mathbb{Z}$, $n \ge 0$, $b_k \in \mathbb{R}$ are constants, and $G_K(x)$ is a function (could be nonlinear) depending on a parameter $K$. 

The $h$-difference Caputo universal $\alpha$-family of maps is defined (derived) as
{\setlength\arraycolsep{0.5pt}   
\begin{eqnarray} 
&&x_{n+1} =   \sum^{m-1}_{k=0}\frac{c_k}{k!}((n+1)h)^{(k)}_h 
\nonumber \\
&&-\frac{h^{\alpha}}{\Gamma(\alpha)}  
\sum^{n+1-m}_{s=0}(n-s-m+\alpha)^{(\alpha-1)} 
G_K(x_{s+m-1}), 
\label{FalFacMap_h}
\end{eqnarray}
}
where $x_k=x(kh)$, $\alpha \in \mathbb{R}$, $\alpha \ge 0$, $m=\lceil \alpha \rceil$, $n \in \mathbb{Z}$, $n \ge m-1$, $c_k \in \mathbb{R}$ are constants, and $G_K(x)$ is a function with a parameter $K$. In both maps 
$h \in \mathbb{R}$ and $h>0$.  
The definition of the falling factorial $t^{(\alpha)}$ is
\begin{equation}
t^{(\alpha)} =\frac{\Gamma(t+1)}{\Gamma(t+1-\alpha)}, \ \ t\ne -1, -2, -3.
...
\label{FrFac}
\end{equation}
The falling factorial is asymptotically a power function:
\begin{equation}
\lim_{t \rightarrow
  \infty}\frac{\Gamma(t+1)}{\Gamma(t+1-\alpha)t^{\alpha}}=1,  
\ \ \ \alpha \in  \mathbb{R}.
\label{GammaLimit}
\end{equation}
The $h$-falling factorial $t^{(\alpha)}_h$ is defined as
\begin{equation}
t^{(\alpha)}_h =h^{\alpha}\frac{\Gamma(\frac{t}{h}+1)}{\Gamma(\frac{t}{h}+1-\alpha)}= h^{\alpha}\Bigl(\frac{t}{h}\Bigr)^{(\alpha)}, \ \ \frac{t}{h} \ne -1, -2, -3,
....
\label{hFrFac}
\end{equation}

In many papers, where particular forms of the Caputo universal map (the Caputo logistic, with $G_K(x)= x-Kx(1-x)$, and the standard, with $G_K(x)= K \sin(x)$, maps) are investigated, the authors assume $h=1$.

\section{Asymptotically periodic cycles for $0<\alpha<1$}
\label{sec:3}

When $0<\alpha<1$, all forms of the universal $\alpha$-family of maps 
introduced in this paper, 
Eqs.~(\ref{FrCMapx})~and~(\ref{FalFacMap_h}), 
can be written in the form
\begin{eqnarray}
x_{n}= x_0 
-\sum^{n-1}_{k=0} G^0(x_k) U(n-k).
\label{FrUUMap}
\end{eqnarray} 

In this formula $G^0(x)=h^\alpha G_K(x)/\Gamma(\alpha)$ and $x_0$ is the initial condition.
In fractional maps Eq.~(\ref{FrCMapx})
\begin{eqnarray}
U_{\alpha}(n)=n^{\alpha-1}, \ \ \ \  U_{\alpha}(1)=1
\label{UnFr}
\end{eqnarray} 
and in fractional difference maps, Eq.~(\ref{FalFacMap_h}),
{\setlength\arraycolsep{0.5pt}
\begin{eqnarray}
&&U_{\alpha}(n)=(n+\alpha-2)^{(\alpha-1)} 
, \ \ \ \  \nonumber \\  
&&U_{\alpha}(1)=(\alpha-1)^{(\alpha-1)}=\Gamma(\alpha).
\label{UnFrDif}
\end{eqnarray} 
}

For $n=lN+m$, where $0<m<l+1$, Eq.~(\ref{FrUUMap}) can be written as
{\setlength\arraycolsep{0.5pt}   
\begin{eqnarray} 
&&x_{lN+m}= x_0 
-\sum^{lN+m-1}_{k=0} G^0(x_k) U_{\alpha} (lN+m-k)
\nonumber \\
&&= x_0 
-\sum^{lN+m}_{n=1} G^0(x_{lN+m-n})U_{\alpha} (n)
 \nonumber \\
&&= x_0 -\sum^{l}_{j=1}\sum^{N-1}_{k=0} G^0(x_{lN+m-lk-j})U_{\alpha} (lk+j)
\nonumber \\
&&-\sum^{m}_{j=1} G^0(x_{m-j}) U_{\alpha} (lN+j).
\label{mth_lCycle_point}
\end{eqnarray}
}
For $0<m<l$
{\setlength\arraycolsep{0.5pt}   
\begin{eqnarray} 
&&x_{lN+m+1}-x_{lN+m}
\nonumber \\
=&&-\sum^{l}_{j=1}\sum^{N-1}_{k=0} G^0(x_{lN+m-lk-j+1})U_{\alpha} (lk+j)
\nonumber \\
&&+\sum^{l}_{j=1}\sum^{N-1}_{k=0} G^0(x_{lN+m-lk-j})U_{\alpha} (lk+j)
\nonumber \\
&&-\sum^{m+1}_{j=1} G^0(x_{m-j+1}) U_{\alpha} (lN+j)+ \sum^{m}_{j=1} G^0(x_{m-j}) U_{\alpha} (lN+j)
\nonumber \\
&&=\sum^{N-1}_{k=0}\Bigl[\sum^{l}_{j=1} G^0(x_{lN+m-lk-j})U_{\alpha} (lk+j)
\nonumber \\
&&-\sum^{l-1}_{j=0} G^0(x_{lN+m-lk-j})U_{\alpha} (lk+j+1)\Bigr]+S_0
\nonumber \\
&&=\sum^{N-1}_{k=0}\sum^{l-1}_{j=1} G^0(x_{lN+m-lk-j})\Bigl[
U_{\alpha} (lk+j)- U_{\alpha} (lk+j+1)\Bigr]
\nonumber \\
&&+\sum^{N-1}_{k=1} G^0(x_{lN+m-lk})\Bigl[
U_{\alpha} (lk)- U_{\alpha} (lk+1)\Bigr]
\nonumber \\
&&- G^0(x_{lN+m})U_{\alpha} (1)+ G^0(x_{m})U_{\alpha} (lN) +S_0,
\label{mth_lCycle_point_dif}
\end{eqnarray}
}
where $S_0$ is a sum of a finite number of elements, which tend to zero as $N \rightarrow \infty$.  Let's assume that in the limit $N \rightarrow \infty$ the system converges to a period $l$ sink ($l$-cycle)
 \begin{equation}
x_{lim,m}=\lim_{N \rightarrow
  \infty} x_{Nl+m}, \  \  \ 0<m<l+1,
\label{TlpointEx}
\end{equation}
and consider the limit of Eq.~(\ref{mth_lCycle_point_dif}) as  
$N \rightarrow \infty$.
{\setlength\arraycolsep{0.5pt}   
\begin{eqnarray} 
&&x_{lim,m+1}-x_{lim,m}
\nonumber \\
&&=\lim_{N \rightarrow \infty}\sum^{l-1}_{j=1}\sum^{N-1}_{k=0}
G^0(x_{lN-lk+m-j})\Bigl[
U_{\alpha} (lk+j)
\nonumber \\
&&- U_{\alpha} (lk+j+1)\Bigr]
+\lim_{N \rightarrow \infty} \sum^{N-1}_{k=1} G^0(x_{lN-lk+m})\Bigl[
U_{\alpha} (lk) 
\nonumber \\
&&- U_{\alpha} (lk+1)\Bigr] - G^0(x_{lim,m})U_{\alpha} (1).
\label{mth_lCycle_point_dif_lim}
\end{eqnarray}
}
Let's find the limit ($0 \le j <l$ and for $j=0$ the sum will start from $k=1$)
{\setlength\arraycolsep{0.5pt}   
\begin{eqnarray} 
&&\lim_{N \rightarrow \infty}\sum^{N-1}_{k=0,1}
G^0(x_{lN-lk+m-j})\Bigl[
U_{\alpha} (lk+j) - U_{\alpha} (lk+j+1)\Bigr]
\nonumber \\
&&=\lim_{N_1 \rightarrow \infty,N-N_1 \rightarrow \infty}\sum^{N_1}_{k=0,1}
G^0(x_{lN-lk+m-j})\Bigl[
U_{\alpha} (lk+j) 
\nonumber \\
&&- U_{\alpha} (lk+j+1)\Bigr]+\lim_{N_1,N-N_1 \rightarrow \infty}\sum^{N}_{k=N_1+1}
G^0(x_{lN-lk+m-j})
\nonumber \\
&&\times \Bigl[
U_{\alpha} (lk+j) - U_{\alpha} (lk+j+1)\Bigr].
\label{TERMlim}
\end{eqnarray}
}

Let's consider the second sum in the last expression. According to our assumption, $x_n$ is converging and it must be bounded. If we assume that $ G^0(x_n)$ is also bounded on a bounded domain, then: 
$|G^0(x_{lN-lk+m-j})|<C_1$. 
The terms $U_{\alpha}(lk+j)-U_{\alpha}(lk+j+1)$ are of the order $(lk+j)^{\alpha-2}$ and the series 
{\setlength\arraycolsep{0.5pt}   
\begin{eqnarray} 
&&S_{j+1}=\sum^{\infty}_{k=0}\Bigl[
U_{\alpha} (lk+j) - U_{\alpha} (lk+j+1)\Bigr], \  \ 0<j<l,
\nonumber \\
&&\tilde{S}_{1}=\sum^{\infty}_{k=1}\Bigl[
U_{\alpha} (lk) - U_{\alpha} (lk+1)\Bigr]
\label{Ser}
\end{eqnarray}
}
are converging. This implies that for every $\varepsilon>0$ there exists a $N_l$ such that 
\begin{equation}
\sum^{N}_{k=N_1+1}\Bigl[U_{\alpha} (lk+j) 
- U_{\alpha} (lk+j+1)\Bigr] < \frac{\varepsilon}{2C_1} 
\label{Ser1}
\end{equation}
for every $N_1>N_l$ and every $N>N_1$ and the limit of this sum when $N_1 \rightarrow \infty$   is zero.

Now, let's consider the first sum of the last expression in Eq.~(\ref{TERMlim}). Because $ G^0(x_{lN-lk+m-j})$ is converging for $N>>N_1$  to $ G^0(x_{lim,m-j})$ for $0 \le j < m$ or to $ G^0(x_{lim,m-j+l})$ for $m \le j < l$, for every $\varepsilon>0$ there exists a large $N$ such that $ G^0(x_{lN-lk+m-j}) = G^0(x_{lim,m-j}) + \varepsilon(k)$ or $ G^0(x_{lN-lk+m-j}) \\ = G^0(x_{lim,m-j+l}) + \varepsilon(k)$, where $|\varepsilon(k)|< \frac{\varepsilon}{2S_{j+1}}$, for every $k \le N_1$. This implies that for every $N_1$ and $\varepsilon>0$ there exists a $N$ such that the considered sum deviates less than $\varepsilon/2$ and the total expression Eq.~(\ref{TERMlim}) deviates less than $\varepsilon$ from 
{\setlength\arraycolsep{0.5pt}   
\begin{eqnarray} 
&& G^0(x_{lim,m-j})S_{j+1},    \   \ 0 < j < m,  
\nonumber \\
&& G^0(x_{lim,m-j+l})S_{j+1},  \   \  m \le j < l, 
\nonumber \\
&& G^0(x_{lim,m})\tilde{S}_{1}, \   \ j=0. 
\label{Limits}
\end{eqnarray}
}
This proves that expressions Eq.~(\ref{Limits}) are the limits as 
$N \rightarrow \infty$ in Eq.~(\ref{TERMlim}).

Finally, Eq.(\ref{mth_lCycle_point_dif_lim}) can be written as 
{\setlength\arraycolsep{0.5pt}   
\begin{eqnarray} 
&&x_{lim,m+1}-x_{lim,m}=S_1 G^0(x_{lim,m})+\sum^{m-1}_{j=1}S_{j+1} G^0(x_{lim,m-j})
\nonumber \\
&&+\sum^{l-1}_{j=m}S_{j+1} G^0(x_{lim,m-j+l}), \  \ 0<m<l,
\label{LimDifferences}
\end{eqnarray}
}
where 
\begin{equation}
{S}_{1}=-U_{\alpha}(1) + \sum^{\infty}_{k=1}\Bigl[
U_{\alpha} (lk) - U_{\alpha} (lk+1)\Bigr].
\label{S1}
\end{equation}
It is easy to see that 
\begin{equation}
\sum^{l}_{j=1}S_j=0.
\label{Ssum}
\end{equation}

To complete the system of equations that defines $l$ variables $x_{lim,j}$, $1 \le j \le l$, we have to add one more equation to the system of $l-1$ equations Eq.~(\ref{LimDifferences}). Let's consider the 
total of all l-cycle limiting points, which is the limiting value of the following sum: 
{\setlength\arraycolsep{0.5pt}   
\begin{eqnarray} 
&& \sum^{l}_{m=1}x_{lN+m}=lx_0
\nonumber \\
&& -\sum^{l}_{m=1}\sum^{l}_{j=1}\sum^{N-1}_{k=0} G^0(x_{lN+m-lk-j})U_{\alpha} (lk+j)
\nonumber \\
&&-\sum^{l}_{m=1}\sum^{m}_{j=1} G^0(x_{m-j}) U_{\alpha} (lN+j).
\label{lCycle_pointTotal}
\end{eqnarray}
}
When $N \rightarrow \infty$, the last term in this equation goes to zero and the terms on the first line of this equation are also finite. This implies that the limiting value of the sum on the second line 
{\setlength\arraycolsep{0.5pt}   
\begin{eqnarray} 
&& S=\sum^{N-N_1}_{k=0}\sum^{l}_{j=1}  \Bigl[ U_{\alpha} (lk+j)\sum^{l}_{m=1} G^0(x_{lN+m-lk-j})\Bigr]
\nonumber \\
&&+\sum^{N}_{k=N-N_1+1}\sum^{l}_{j=1}  \Bigl[ U_{\alpha} (lk+j)\sum^{l}_{m=1} G^0(x_{lN+m-lk-j})\Bigr]
\label{FiniteSum}
\end{eqnarray}
}
also must be finite. For a finite value of $N_1$, the second sum in the last expression is finite. For large values of $N-k$, $\sum^{l}_{m=1} G^0(x_{lN+m-lk-j})$ is converging to some constant value $S_G$.
\begin{equation}
S_G=\lim_{N-k \rightarrow \infty} \sum^{l}_{j=1} G^0(x_{lN+m-lk-j}). 
\label{SG}
\end{equation}
If $S_G > 0$, then there exists a $N_1$ such that for every $k$ in the first sum $ G^0(x_{lN+m-lk-j})>C_1>0$; if $S_G < 0$ then there exists a $N_1$ such that for every $k$ in the first sum $G^0(x_{lN+m-lk-j})<C_2<0$. In both cases the first sum is diverging when $N \rightarrow \infty$. The only case in which the limiting value of $S$ is finite is when $S_G=0$:
\begin{equation}
\sum^{l}_{j=1} G^0(x_{lim,j})=0.
\label{Close}
\end{equation}
The system of $l$ equations, Eq.~(\ref{LimDifferences}) and  Eq.~(\ref{Close}), is a system that defines all $l$ values $x_{lim,m}$,
$1 \le m \le l$ of an asymptotic $l$-cycle.

\section{Calculation of sums for the $T=l$-cycles}
\label{sec:4}

The first step in solving equations Eqs.~(\ref{LimDifferences})~and~(\ref{Close}), which define periodic points, is to compute the sums $S_j$ defined by Eqs.~(\ref{Ser})~and~(\ref{S1}). The order of the terms in those sums is $k^{\alpha-2}$ and they converge very slowly. 
 
\subsection{Fractional maps}
\label{sec:4a}

In factional maps, the functions $U_\alpha (n)$ are defined by Eq.~(\ref{UnFr}). For $S_1$ we may write
\begin{equation}
{S}_{1}=S_{1,1}+S_{1,2},
\label{S1nl}
\end{equation}
where the finite series
\begin{equation}
{S}_{1,1}=-1 + \sum^{N}_{k=1}\Bigl[
(lk)^{\alpha-1} - (lk+1)^{\alpha-1} \Bigr]
\label{S1n1l}
\end{equation}
with a large $N$ (to tabulate values of $S_i$ we used $N=20000$) can be directly calculated with a high (machine) accuracy. To calculate the infinite series
\begin{equation}
{S}_{1,2}=\sum^{\infty}_{k=N+1}\Bigl[
(lk)^{\alpha-1} - (lk+1)^{\alpha-1} \Bigr]
\label{S1n2l}
\end{equation}
in each term we factor out $(lk)^{\alpha-1}$ and develop the difference into a Taylor series. At the end, the expression to calculate ${S}_{1,2}$ can be written as
{\setlength\arraycolsep{0.5pt}   
\begin{eqnarray} 
&&{S}_{1,2}=(1-\alpha)l^{\alpha-2}\Biggl\{\zeta_N(2-\alpha) +\frac{\alpha-2}{2l}\Biggl[\zeta_N(3-\alpha)
\nonumber \\
&& +\frac{\alpha-3}{3l}\Biggl(\zeta_N(4-\alpha)+\frac{\alpha-4}{4l}\zeta_N(5-\alpha)\Biggr)\Biggr]\Biggr\}
\nonumber \\
&&+O(N^{\alpha-5}),
\label{S1nfl}
\end{eqnarray}
}
where 
\begin{equation}
\zeta_N(m-\alpha)=\zeta(m-\alpha)-\sum^{N}_{k=1}k^{\alpha-m},
\label{ZetaNl}
\end{equation}
and we used a fast method for calculating values of the Riemann $\zeta$-function.
In a similar way, the expressions for $S_{j+1}$ for $0<j<l$ can be written as 
{\setlength\arraycolsep{0.5pt}   
\begin{eqnarray} 
&&{S}_{j+1}=\sum^{N}_{k=0}\Bigl[
(lk+j)^{\alpha-1} - (lk+j+1)^ {\alpha-1} \Bigr]
\nonumber \\
&&+(1-\alpha)l^{\alpha-2}\Biggl\{\zeta_N(2-\alpha) +\frac{\alpha-2}{2l}\Biggl[(2j+1)\zeta_N(3-\alpha)
\nonumber \\
&& +\frac{\alpha-3}{3l}\Biggl((3j^2+3j+1)\zeta_N(4-\alpha)+\frac{(\alpha-4)}{4l}(2j+1)
\nonumber \\
&&\times (2j^2+2j+1)\zeta_N(5-\alpha)\Biggr)\Biggr]\Biggr\}
+O(N^{\alpha-5}).
\label{Sjnl}
\end{eqnarray}
}

\subsection{Fractional difference maps}
\label{sec:4b}

In factional difference maps the functions $U_\alpha (n)$ are defined by Eq.~(\ref{UnFrDif}). In this case, each term of the sums in 
Eq.~(\ref{Ser}) can be written as 
{\setlength\arraycolsep{0.5pt}   
\begin{eqnarray} 
&&U_\alpha(lk+j)-U_\alpha(lk+j+1)=\frac{\Gamma(lk+j+\alpha-1)}{\Gamma(lk+j)} 
\nonumber \\
&&- \frac{\Gamma(lk+j+\alpha+)}{\Gamma(lk+j+1)}
= \frac{(1-\alpha)*\Gamma(lk+j+\alpha)-1}{\Gamma(lk+j+1)}.
\label{termFDl}
\end{eqnarray}
}
As in the fractional case, we will split the total into two sums:
\begin{equation}
{S}_{j+1}=S_{j+1,1}+S_{j+1,2},
\label{S1nFD}
\end{equation}
where the finite series
\begin{equation}
{S}_{j+1,1}=\sum^{N}_{k=0}\Biggl[\frac{(1-\alpha)\Gamma(lk+j+\alpha-1)}{(lk+j)!}
 \Biggr]
\label{S2n1l}
\end{equation}
can be directly calculated with a high accuracy. In the fractional difference case with $j=0$, the $k=0$ term is equal to $-U_{\alpha}(1)=-\Gamma(\alpha)$ and Eq.~(\ref{S2n1l}) defines the sums ${S}_{j+1,1}$ for all $l$ values of $j$: $0 \le j <l$. 
To calculate the infinite series
\begin{equation}
{S}_{j+1,2}=(1-\alpha)\sum^{\infty}_{k=N+1}\Biggl[\frac{\Gamma(lk+j+\alpha-1)}{\Gamma(lk+j+1)}
 \Biggr]
\label{S2n2l}
\end{equation}
we will use the following approximation (\cite{GammaRat}):
{\setlength\arraycolsep{0.5pt}   
\begin{eqnarray} 
&&\frac{\Gamma(z+a)}{\Gamma(z+b)}=z^{a-b}\Biggl\{1+ \frac{(a+b-1)(a-b)}{2z}+\frac{1}{12z^2}\left( \begin{array}{c} a-b\\ 2
\end{array} \right)
\nonumber \\
&& \times 
[3(a+b-1)^2-(a-b+1)]+O(z^{-3})\Biggr\}.
\label{GammaRat}
\end{eqnarray}
}
Then, we obtain the following expression to calculate $S_{j+1}$:
{\setlength\arraycolsep{0.5pt}   
\begin{eqnarray} 
&&S_{j+1}=(1-\alpha) \sum^{N}_{k=0}\Biggl[\frac{\Gamma(lk+j+\alpha-1)}{(lk+j)!}
 \Biggr]
\nonumber \\
&&+(1-\alpha)l^{\alpha-2}\Biggl\{\zeta_N(2-\alpha) +\frac{\alpha-2}{2l}\Biggl[(2j+\alpha-1)\zeta_N(3-\alpha)
\nonumber \\
&& +\frac{(\alpha-3)[3(2j+\alpha-1)^2-\alpha+1]}{12l}\zeta_N(4-\alpha)\Biggr]\Biggr\}
\nonumber \\
&&+O(N^{\alpha-4}),   \ \ \  0 \le j <l.
\label{S2FrDifLastl}
\end{eqnarray}
}

\section{The sums $S_j$ for period 3 cycles}
\label{sec:5}

In this section we present the expressions for the $l=3$-cycle sums $S_1$, $S_2$, and $S_3$. 

\subsection{Fractional maps}
\label{sec:5.1}

For fractional maps Eqs.~(\ref{S1nl})-(\ref{S1nfl}) can be written in the form
{\setlength\arraycolsep{0.5pt}   
\begin{eqnarray} 
&&{S}_1=-1 + \sum^{N}_{k=1}\Bigl[
(3k)^{\alpha-1} - (3k+1)^ {\alpha-1} \Bigr]
\nonumber \\
&&(1-\alpha)3^{\alpha-2}\Biggl\{\zeta_N(2-\alpha) +\frac{\alpha-2}{6}\Biggl[\zeta_N(3-\alpha)
\nonumber \\
&& +\frac{\alpha-3}{9}\Biggl(\zeta_N(4-\alpha)+\frac{\alpha-4}{12}\zeta_N(5-\alpha)\Biggr)\Biggr]\Biggr\}
\nonumber \\
&&+O(N^{\alpha-5}).
\label{S1nf}
\end{eqnarray}
}
The expressions for $S_2$ and $S_3$ obtained from Eq.~(\ref{Sjnl}) can be written as 
{\setlength\arraycolsep{0.5pt}   
\begin{eqnarray} 
&&{S}_2=\sum^{N}_{k=0}\Bigl[
(3k+1)^{\alpha-1} - (3k+2)^ {\alpha-1} \Bigr]
\nonumber \\
&&+(1-\alpha)3^{\alpha-2}\Biggl\{\zeta_N(2-\alpha) +\frac{\alpha-2}{2}\Biggl[\zeta_N(3-\alpha)
\nonumber \\
&& +\frac{\alpha-3}{27}\Biggl(7\zeta_N(4-\alpha)+\frac{5(\alpha-4)}{4}\zeta_N(5-\alpha)\Biggr)\Biggr]\Biggr\}
\nonumber \\
&&+O(N^{\alpha-5})
\label{S2n}
\end{eqnarray}
}
and
{\setlength\arraycolsep{0.5pt}   
\begin{eqnarray} 
&&{S}_3=\sum^{N}_{k=0}\Bigl[
(3k+2)^{\alpha-1} - (3k+3)^ {\alpha-1} \Bigr]
\nonumber \\
&&+(1-\alpha)3^{\alpha-2}\Biggl\{\zeta_N(2-\alpha) +\frac{\alpha-2}{6}\Biggl[5\zeta_N(3-\alpha)
\nonumber \\
&& +\frac{\alpha-3}{9}\Biggl(19\zeta_N(4-\alpha)+\frac{65(\alpha-4)}{12}\zeta_N(5-\alpha)\Biggr)\Biggr]\Biggr\}
\nonumber \\
&&+O(N^{\alpha-5})
\label{S3n}
\end{eqnarray}
}
To calculate values of $S_i$ in this paper (see Tables~\ref{table:T3_Fr}~and~\ref{table:T3_FrDif}) we used $N=20000$ and a fast method to calculate the $\zeta$-function. To estimate the accuracy of our computations, we also calculated the value of $\Sigma S_i$, whose deviation from zero represents the absolute error.

\begin{table}[ht!]
\centering
    \begin{tabular}{| c  | c       |  c |    c  | c   | c   | c       |}
    \hline
    $\alpha$ & $S_1$ & $S_2$ & $S_3$  &  $\Sigma S_i$    \\ \hline
    0.01  & -.8503346 & .6023682 & .2479663 & 6.2e-14 \\ \hline
    0.05  & -.8451510 & .5933434 & .2518076 & 6.2e-14 \\ \hline
    0.1   & -.8384473 & .5818399 & .2566074 & 6.4e-14 \\ \hline
    0.15  & -.8314875 & .5700882 & .2613993 & 5.4e-14 \\ \hline
    0.2   & -.8242640 & .5580876 & .2661764 & 3.2e-14 \\ \hline
    0.25  & -.8167694 & .5458379 & .2709315 & 1.5e-15 \\ \hline
    0.3   & -.8089959 & .5333390 & .2756569 & -5.5e-14 \\ \hline
    0.35  & -.8009359 & .5205915 & .2803445 & -6.7e-16 \\ \hline
    0.4   & -.7925818 & .5075961 & .2849857 & -2.4e-14 \\ \hline
    0.45  & -.7839259 & .4943542 & .2895718 & -5.8e-15 \\ \hline
    0.5   & -.7749606 & .4808676 & .2940931 & -1.0e-14 \\ \hline
    0.55  & -.7656783 & .4671385 & .2985398 & -2.0e-14 \\ \hline
    0.6   & -.7560713 & .4531697 & .3029016 & -2.1e-15 \\ \hline
    0.65  & -.7461323 & .4389647 & .3071676 & -1.8e-14 \\ \hline
    0.7   & -.7358540 & .4245274 & .3113265 & 1.3e-15 \\ \hline
    0.75  & -.7252290 & .4098625 & .3153665 & -1.1e-14 \\ \hline
    0.8   & -.7142504 & .3949752 & .3192752 & -1.0e-14 \\ \hline
    0.85  & -.7029113 & .3798715 & .3230398 & 1.3e-14 \\ \hline
    0.9   & -.6912052 & .3645582 & .3266470 & -2.7e-14 \\ \hline
    0.95  & -.6791256 & .3490427 & .3300830 & 5.0e-14 \\ \hline
    0.99  & -.6691891 & .3364903 & .3326988 & 2.1e-13 \\ \hline
    \end{tabular}
    \caption{The values of $S_j$, $1 \le j \le 3$, for period 3 cycles (fractional maps).}
    \label{table:T3_Fr}
\end{table}

\subsection{Fractional difference maps}
\label{sec:5.2}
In factional difference maps Eq.~(\ref{S2FrDifLastl}) produces the following expressions for $S_1$, $S_2$, and $S_3$:
{\setlength\arraycolsep{0.5pt}   
\begin{eqnarray} 
&&{S}_1=-\Gamma(\alpha)+ (1-\alpha)\sum^{N}_{k=1}\Biggl[\frac{\Gamma(3k+\alpha-1)}{(3k)!}
 \Biggr]
\nonumber \\
&&+(1-\alpha)3^{\alpha-2}\Biggl\{\zeta_N(2-\alpha) +\frac{\alpha-2}{6}\Biggl[(\alpha-1)\zeta_N(3-\alpha)
\nonumber \\
&& +\frac{(\alpha-3)(3\alpha^2-7\alpha+4)}{36}\zeta_N(4-\alpha)\Biggr]\Biggr\}
+O(N^{\alpha-4}),
\label{S1FrDifLast}
\end{eqnarray}
}
{\setlength\arraycolsep{0.5pt}   
\begin{eqnarray} 
&&{S}_2=(1-\alpha)\sum^{N}_{k=0}\Biggl[\frac{\Gamma(3k+\alpha)}{(3k+1)!}
 \Biggr]
\nonumber \\
&&+(1-\alpha)3^{\alpha-2}\Biggl\{\zeta_N(2-\alpha) +\frac{\alpha-2}{6}\Biggl[(\alpha+1)\zeta_N(3-\alpha)
\nonumber \\
&& +\frac{(\alpha-3)(3\alpha^2+5\alpha+4)}{36}\zeta_N(4-\alpha)\Biggr]\Biggr\}
+O(N^{\alpha-4}),
\label{S2FrDifLast}
\end{eqnarray}
}
and
{\setlength\arraycolsep{0.5pt}   
\begin{eqnarray} 
&&{S}_3=(1-\alpha)\sum^{N}_{k=0}\Biggl[\frac{\Gamma(3k+1+\alpha)}{(3k+2)!}
 \Biggr]
\nonumber \\
&&+(1-\alpha)3^{\alpha-2}\Biggl\{\zeta_N(2-\alpha) +\frac{\alpha-2}{6}\Biggl[(\alpha+3)\zeta_N(3-\alpha)
\nonumber \\
&& +\frac{(\alpha-3)(3\alpha^2+17\alpha+28)}{36}\zeta_N(4-\alpha)\Biggr]\Biggr\}
\nonumber \\
&&+O(N^{\alpha-4}).
\label{S3FrDifLast}
\end{eqnarray}
}
The results of calculations are given in Table~\ref{table:T3_FrDif}.

\begin{table}[ht!]
\centering
    \begin{tabular}{| c  | c       |      c  | c      | c       |}
    \hline
    $\alpha$ & $S_1$  & $S_2$    & $S_3$    &  $\Sigma S_i$    \\ \hline
    0.01  & -99.18547 & 98.58760 & .5978733 & 6.3e-13 \\ \hline
    0.05  & -19.22240 & 18.64983 & .5725695 & 1.1e-14 \\ \hline
    0.1   & -9.264775 & 8.720580 & .5441945 & 3.1e-14 \\ \hline 
    0.15  & -5.970119 & 5.451154 & .5189649 & 7.1e-15 \\ \hline
    0.2   & -4.338903 & 3.842443 & .4964601 & 6.9e-14\\ \hline
    0.25  & -3.371519 & 2.895189 & .4763300 & 7.3e-15\\ \hline
    0.3   & -2.734966 & 2.276684 & .4582818 & -4.9e-15\\ \hline
    0.35  & -2.286668 & 1.844600 & .4420686 & 2.6e-15\\ \hline
    0.4   & -1.955439 & 1.527958 & .4274810 & -2.00e-14\\ \hline
    0.45  & -1.701803 & 1.287462 & .4143406 & 1.4e-14\\ \hline
    0.5   & -1.502131 & 1.099636 & .4024948 & 2.8e-14\\ \hline
    0.55  & -1.341428 & .9496159 & .3918119 & 6.1e-16\\ \hline
    0.6   & -1.209729 & .8275508 & .3821784 & 1.2e-14\\ \hline
    0.65  & -1.100162 & .7266661 & .3734958 & -1.0e-14\\ \hline
    0.7   & -1.007836 & .6421581 & .3656784 & 1.8e-14\\ \hline
    0.75  & -.9291830 & .5705318 & .3586513 & -4.1e-15\\ \hline
    0.8   & -.8615369 & .5091881 & .3523488 & -1.1e-14\\ \hline
    0.85  & -.8028708 & .4561572 & .3467136 & -8.4e-15\\ \hline
    0.9   & -.7516160 & .4099212 & .3416948 & 9.4e-16\\ \hline
    0.95  & -.7065412 & .3692933 & .3372479 & 6.3e-14\\ \hline
    0.99  & -.6742658 & .3401904 & .3340754 & 2.0e-13\\ \hline
    \end{tabular}
    \caption{The values of $S_j$, $1 \le j \le 3$, for $T=3$ cycles (fractional difference maps).}
    \label{table:T3_FrDif}
\end{table}

\section{Period 2 cycles}
\label{sec:6}

For $T=2$ cycles the equality $S_1=-S_2$ can be directly verified from the definition of $S_i$ (Eqs.~(\ref{Ser}),~(\ref{S1}),~and~(\ref{Ssum})). 
The analysis of $T=2$ cycles is put here after the analysis of $T=3$ cycles because these cycles were analyzed in the author's previous papers. 

The case of fractional maps was analyzed in \cite{Chaos}. It is easy to see that the variable $V_{\alpha l}$ introduced in that paper is equal to $-2S_1$. The formula to compute $V_{\alpha l}$ (Eq.~(A4) from \cite{Chaos}) can be transformed to compute values of $S_1$:
 {\setlength\arraycolsep{0.5pt}   
\begin{eqnarray} 
&&{S}_1=-1 + \sum^{N-1}_{k=1}\Bigl[
(2k)^{\alpha-1} - (2k+1)^ {\alpha-1} \Bigr]+0.5(2N)^{\alpha-1}
\nonumber \\
&&+2^{\alpha-4}(1-\alpha)(2-\alpha) \Biggl\{\zeta_N(3-\alpha)
+\frac{3-\alpha}{2}\Biggl[\zeta_N(4-\alpha)
\nonumber \\
&&+\frac{4-\alpha}{8}\Biggl(\frac{7}{3}\zeta_N(5-\alpha)+ \frac{5-\alpha}{2}\zeta_N(6-\alpha)\Biggr)\Biggr]\Biggr\}
\nonumber \\
&&+O(N^{\alpha-6}).
\label{S1T2n}
\end{eqnarray}
}
From Eq.~(\ref{S1})
{\setlength\arraycolsep{0.5pt}   
\begin{eqnarray} 
&&{S}_{1}=-1 + \sum^{N}_{k=1}\Bigl[
(2k)^{\alpha-1} - (2k+1)^ {\alpha-1} \Bigr]
\nonumber \\
&& +(1-\alpha)2^{\alpha-2}\Biggl\{\zeta_N(2-\alpha) +\frac{\alpha-2}{4}\Biggl[\zeta_N(3-\alpha)
\nonumber \\
&& +\frac{\alpha-3}{6}\Bigl[\zeta_N(4-\alpha)+\frac{\alpha-4}{8}\zeta_N(5-\alpha)\Bigr]\Biggr]\Biggr\}
\nonumber \\
&&+O(N^{\alpha-5}).
\label{S1T2}
\end{eqnarray}
}

A formula for the calculation of $S_1$ in the case of fractional difference maps has never been published but was used in \cite{Chaos2018,Chaos2014,ME9} to calculate $T=2$ points and bifurcation diagrams in the fractional difference logistic and standard maps:
{\setlength\arraycolsep{0.5pt}   
\begin{eqnarray} 
&&{S}_1=-\Gamma(\alpha)+ (1-\alpha)\sum^{N}_{k=1}\Biggl[\frac{\Gamma(2k+\alpha-1)}{(2k)!}
 \Biggr]
\nonumber \\
&&+(1-\alpha)2^{\alpha-2}\Biggl\{\zeta_N(2-\alpha) +\frac{\alpha-2}{4}\Biggl[(\alpha-1)\zeta_N(3-\alpha)
\nonumber \\
&& +\frac{(\alpha-3)(3\alpha^2-7\alpha+4)}{24}\zeta_N(4-\alpha)\Biggr]\Biggr\}
+O(N^{\alpha-4})
\label{S1FrDifT2}
\end{eqnarray}
}

\begin{table}[ht!]
\centering
    \begin{tabular}{| c  | c       |      c  | c      |}
    \hline
    $\alpha$ & $S_1$ & $S_2$ &  $\Sigma S_i$    \\ \hline
    0.01  & -.6915452& .6915452& 1.0e-13 \\ \hline
    0.05  & -.6850718& .6850718& 9.8e-14 \\ \hline
    0.1   & -.6768319& .6768319& 9.1e-14 \\ \hline
    0.15  & -.6684265& .6684265& 8.3e-14 \\ \hline
    0.2   & -.6598547& .6598547& 4.6e-14 \\ \hline
    0.25  & -.6511157& .6511157& -5.1e-15 \\ \hline
    0.3   & -.6422090& .6422090& -7.1e-14 \\ \hline
    0.35  & -.6331341& .6331341& -4.4e-15 \\ \hline
    0.4   & -.6238908& .6238908& -2.4e-14 \\ \hline
    0.45  & -.6144789& .6144789& -6.1e-15 \\ \hline
    0.5   & -.6048986& .6048986& -2.3e-14 \\ \hline
    0.55  & -.5951502& .5951502& -2.9e-14 \\ \hline
    0.6   & -.5852340& .5852340& -1.4e-14 \\ \hline
    0.65  & -.5751509& .5751509& -1.8e-14 \\ \hline
    0.7   & -.5649016& .5649016& -8.2e-15 \\ \hline
    0.75  & -.5544874& .5544874& -2.0e-14 \\ \hline
    0.8   & -.5439096& .5439096& -2.1e-14 \\ \hline
    0.85  & -.5331700& .5331700& 9.8e-15 \\ \hline
    0.9   & -.5222703& .5222703& -1.9e-14 \\ \hline
    0.95  & -.5112128& .5112128& 5.1e-14 \\ \hline
    0.99  & -.5022549& .5022549& 2.1e-13 \\ \hline
    \end{tabular}
    \caption{The values of $S_j$, $1 \le j \le 2$, for period 2 cycles (fractional maps).}
    \label{table:T2_Fr}
\end{table}

\begin{table}[ht!]
\centering
    \begin{tabular}{| c  | c       |      c  | c      |}
    \hline
    $\alpha$ & $S_1$ & $S_2$ &  $\Sigma S_i$    \\ \hline
    0.01  & -98.74575& 98.74575& 2.3e-13 \\ \hline
    0.05  & -18.80686& 18.80686& -9.2e-14 \\ \hline
    0.1   & -8.876417& 8.876417& 3.9e-14 \\ \hline
    0.15  & -5.606024& 5.606024& 4.2e-14 \\ \hline
    0.2   & -3.996562& 3.996562& -1.8e-18 \\ \hline
    0.25  & -3.048762& 3.048762& -3.6e-15 \\ \hline
    0.3   & -2.429909& 2.429909& -2.4e-14 \\ \hline
    0.35  & -1.997666& 1.997666& 4.7e-15 \\ \hline
    0.4   & -1.681051& 1.681051& -1.0e-14 \\ \hline
    0.45  & -1.440760& 1.440760& -1.3e-15 \\ \hline
    0.5   & -1.253314& 1.253314& 1.5e-14 \\ \hline
    0.55  & -1.103845& 1.103845& -8.9e-15 \\ \hline
    0.6   & -.9825005& .9825005& 1.7e-14 \\ \hline
    0.65  & -.8825027& .8825027& -1.4e-14 \\ \hline
    0.7   & -.7990468& .7990468& 7.5e-15 \\ \hline
    0.75  & -.7286371& .7286371& -3.6e-15 \\ \hline
    0.8   & -.6686744& .6686744& -2.2e-14 \\ \hline
    0.85  & -.6171890& .6171890& 1.2e-14 \\ \hline
    0.9   & -.5726639& .5726639& -1.8e-14 \\ \hline
    0.95  & -.5339137& .5339137& 6.7e-14 \\ \hline
    0.99  & -.5064342& .5064342& 2.1e-13 \\ \hline
    \end{tabular}
    \caption{The values of $S_j$, $1 \le j \le 2$, for T=2 cycles (fractional difference maps).}
    \label{table:T2_FrDif}
\end{table}

\section{Bifurcations and asymptotically periodic points in logistic $\alpha$-families of maps ($0 \le \alpha \le 1$)}
\label{sec:7}

Information on the first bifurcations and cycle 2 sink points in the fractional/fractional difference standard, with $G_K(x)=K \sin(x)$, and logistic, with $G_K(x)=x-Kx(1-x)$, families of maps for $0 \le \alpha \le 2$ can be found in \cite{HBV2,HBV4} and references therein. Here we will consider only the logistic families of maps and present more detailed results.

\subsection{$T=2$ cycles}
\label{sec:7.1}

In the case $T=2$, the system of equations Eq.~(\ref{LimDifferences}) and  Eq.~(\ref{Close}) can be written as
\begin{equation}
\begin{array}{c}
\left\{
\begin{array}{lll}
(1-K)(x_{lim,1}+x_{lim,2})+K(x_{lim,1}^2+x_{lim,2}^2)=0,
\\ 
x_{lim,1} - x_{lim,2} = \frac{S_2}{2\Gamma(\alpha)}
h^{\alpha}  (x_{lim,1}-x_{lim,2})[1
\\-K+(x_{lim,1}+x_{lim,2})].
\end{array}
\right.
\end{array} 
\label{T2Logi}
\end{equation} 
Two fixed-point solutions with $x_{lim,1}=x_{lim,2}$ are $x_{lim,1}=0$, stable for
$K<1$, and $x_{lim,1}=(K-1)/K$.

The $T=2$ sink is defined by the equation 
{\setlength\arraycolsep{0.5pt}   
\begin{eqnarray} 
&&x_{lim,1}^2-\Bigl( \frac{\Gamma(\alpha)}{S_2Kh^{\alpha}} 
+\frac{K-1}{K}  \Bigr)x_{lim,1} 
\nonumber \\
&&+\frac{\Gamma^2(\alpha)}{2(S_2Kh^{\alpha})^2}
+\frac{(K-1)\Gamma(\alpha)}{2S_2K^2h^{\alpha}}=0,
\label{eqT2log}
\end{eqnarray}
}
which has solutions 
\begin{equation}
x_{lim,1}=\frac{K_{C1s}+K-1 \pm \sqrt{(K-1)^2-K_{C1s}^2}}{2K}
\label{eqT2logSolu},
\end{equation}
where (see \cite{Chaos})
\begin{equation}
K_{C1s}=\frac{\Gamma(\alpha)}{S_2 h^{\alpha}}
\label{Kc1s}
\end{equation}
is a first bifurcation point of the standard families of maps,
defined when
\begin{equation}
K \ge 1 + \frac{\Gamma(\alpha)}{S_2h^{\alpha}}=1+K_{C1s} \ \ {\rm or} \ \
K \le 1 - \frac{\Gamma(\alpha)}{S_2h^{\alpha}}=1-K_{C1s}. 
\label{condT2log}
\end{equation}
Conditions Eq.~(\ref{condT2log}) are valid for all 
logistic $\alpha$-families of maps considered in this paper. 
Here we'll consider $K>0$ and $h \le 1$. It is easy to show, and this is done in \cite{Chaos2018}, that $S_2$ is less than $U_{\alpha}(1)/2$, which is either $1/2$ or
$\Gamma(\alpha)/2$ (also $\Gamma(\alpha)>0.885$ for $\alpha>0$).
Then, $ \Gamma(\alpha)/(S_2 h^{\alpha})>1$ and we may ignore the
second of the inequalities in Eq.~(\ref{condT2log}). Note that
the fixed point $x=(K-1)/K$ is stable when
\begin{equation}
1 \le K <K_{C1l}= 1 + \frac{\Gamma(\alpha)}{S_2h^{\alpha}}=1+K_{C1s}.
\label{condT2logn}
\end{equation} 

\begin{figure}[!t]
\includegraphics[width=0.45 \textwidth]{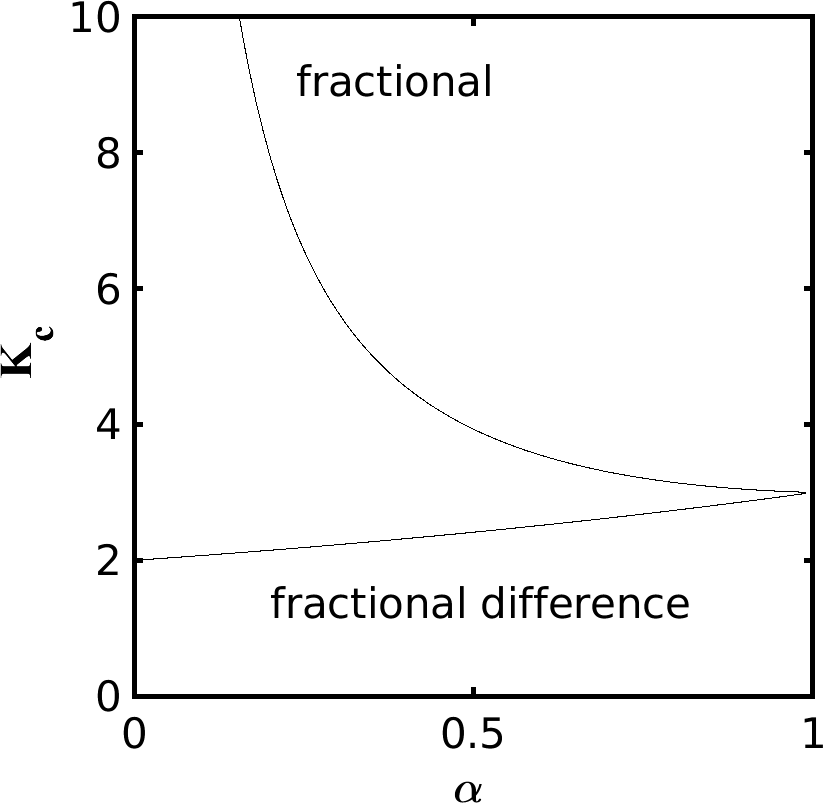}
\vspace{-0.25cm}
\caption{Bifurcation K-$\alpha$ curves on which transition from a fixed point to a $T=2$ cycle occurs for fractional (upper curve) and fractional difference logistic maps ($h=1$).
}
\label{fig1}
\end{figure}

Below we present the table (Table~\ref{table:T2_FrDifBif}) of the 
fixed-point -- $T=2$-cycle bifurcation points for fractional and fractional difference maps and the corresponding graph, Fig.~\ref{fig1}, which is a part of the previously reported \cite{HBV4} 2D bifurcation diagram, where we assume $h=1$.

\begin{table}[ht!]
\centering
    \begin{tabular}{| c  | c       |      c  | }
    \hline
    $\alpha$ & Fractional & Fr. difference      \\ \hline
    0.01  & 144.7832& 2.006956 \\ \hline
    0.05  & 29.42050& 2.035265 \\ \hline
    0.1   & 15.05594& 2.071773 \\ \hline
    0.15  & 10.30584& 2.109569 \\ \hline
    0.2   & 7.957356& 2.148698 \\ \hline
    0.25  & 6.568304& 2.189207 \\ \hline
    0.3   & 5.658249& 2.231144 \\ \hline
    0.35  & 5.021497& 2.274561 \\ \hline
    0.4   & 4.555365& 2.319508 \\ \hline
    0.45  & 4.202936& 2.366040 \\ \hline
    0.5   & 3.930167& 2.414214 \\ \hline
    0.55  & 3.715490& 2.464086 \\ \hline
    0.6   & 3.544610& 2.515717 \\ \hline
    0.65  & 3.407708& 2.569168 \\ \hline
    0.7   & 3.297843& 2.624505 \\ \hline
    0.75  & 3.209999& 2.681793 \\ \hline
    0.8   & 3.140484& 2.741101 \\ \hline
    0.85  & 3.086546& 2.802501 \\ \hline
    0.9   & 3.046122& 2.866066 \\ \hline
    0.95  & 3.017659& 2.931873 \\ \hline
    0.99  & 3.002712& 2.986185 \\ \hline
    \end{tabular}
    \caption{Fixed-point -- $T=2$-cycle bifurcation points for fractional (middle column) and fractional difference (right column) maps ($h=1$). }
    \label{table:T2_FrDifBif}
\end{table}

\begin{figure}[!t]
\includegraphics[width=0.45 \textwidth]{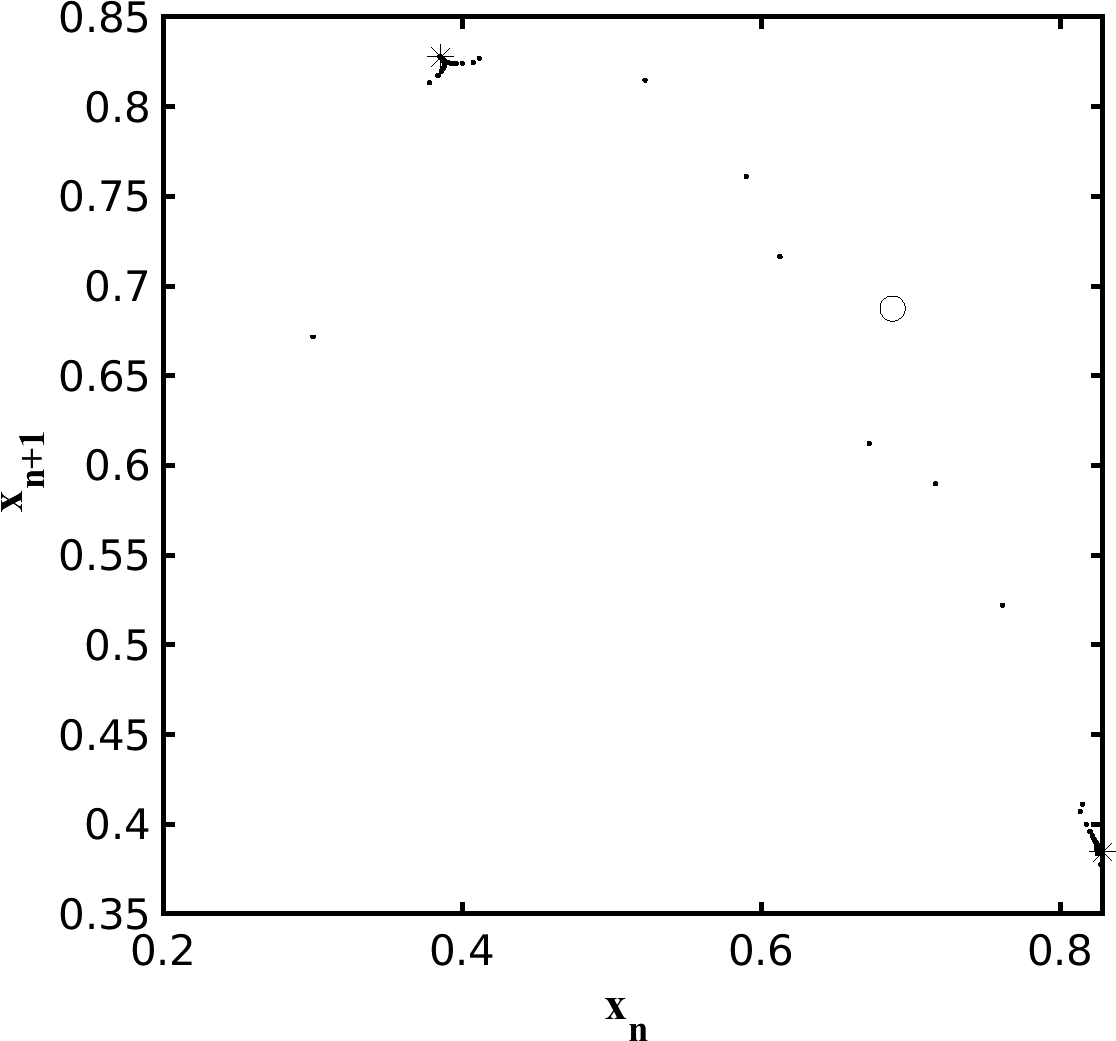}
\vspace{-0.25cm}
\caption{ The Poincar\'{e} plot (500000 iterations) for fractional difference logistic map with $\alpha=0.75$, $K=3.2$, $h=1$, and the initial point $x_0=0.3$. The asymptotically stable $T=2$ sink is marked by the stars and the unstable fixed point $(K-1)/K$ by the circle.
}
\label{fig2}
\end{figure}

\begin{figure}[!t]
\includegraphics[width=0.45 \textwidth]{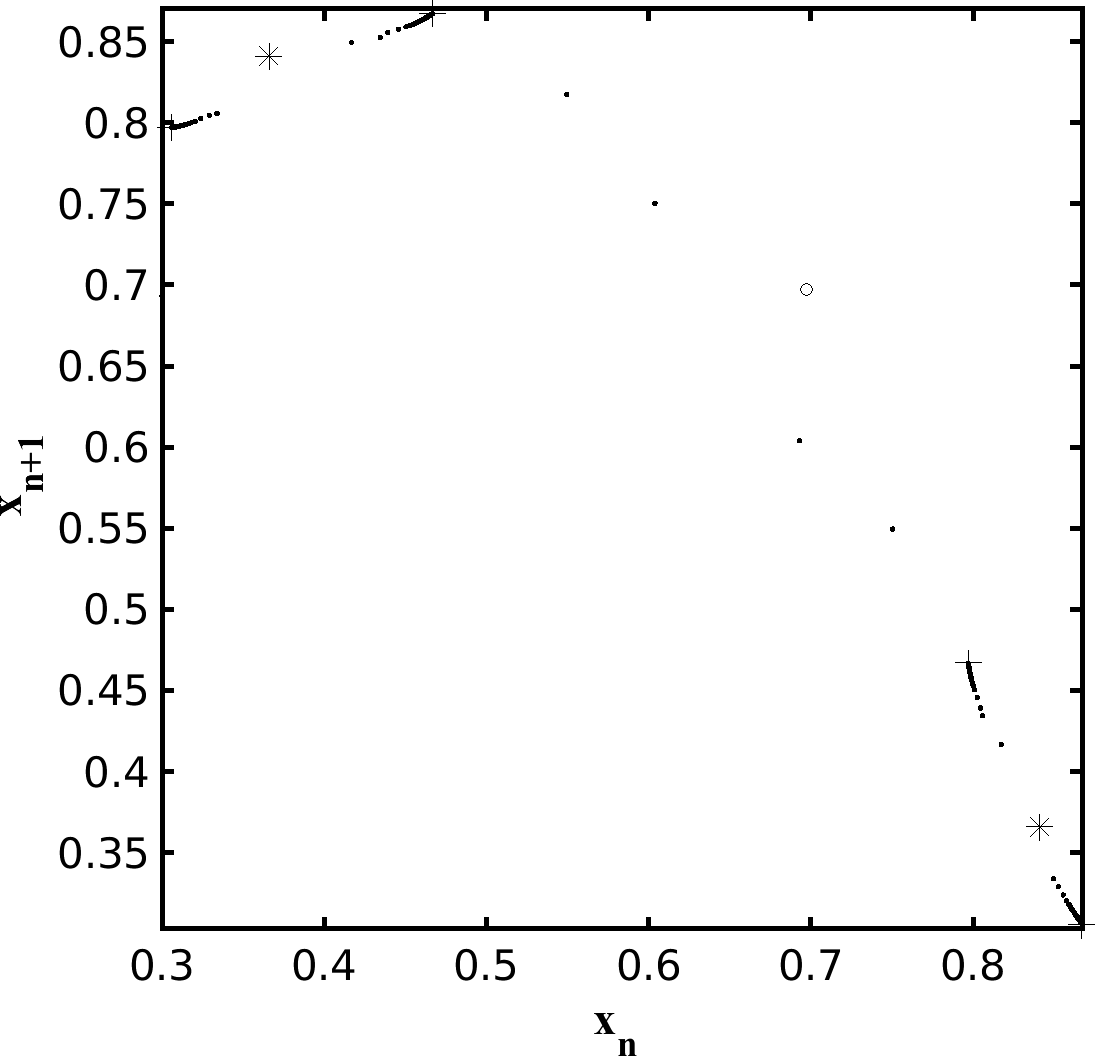}
\vspace{-0.25cm}
\caption{ The Poincar\'{e} plot (500000 iterations) for fractional difference logistic map with $\alpha=0.75$, $K=3.3$, $h=1$, and the initial point $x_0=0.3$. Stable $T=4$ sink marked by the plus signs. The asymptotically unstable $T=2$ sink is marked by the stars and the unstable fixed point $(K-1)/K$ by the circle.
}
\label{fig3}
\end{figure} 

\begin{figure}[!t]
\includegraphics[width=0.45 \textwidth]{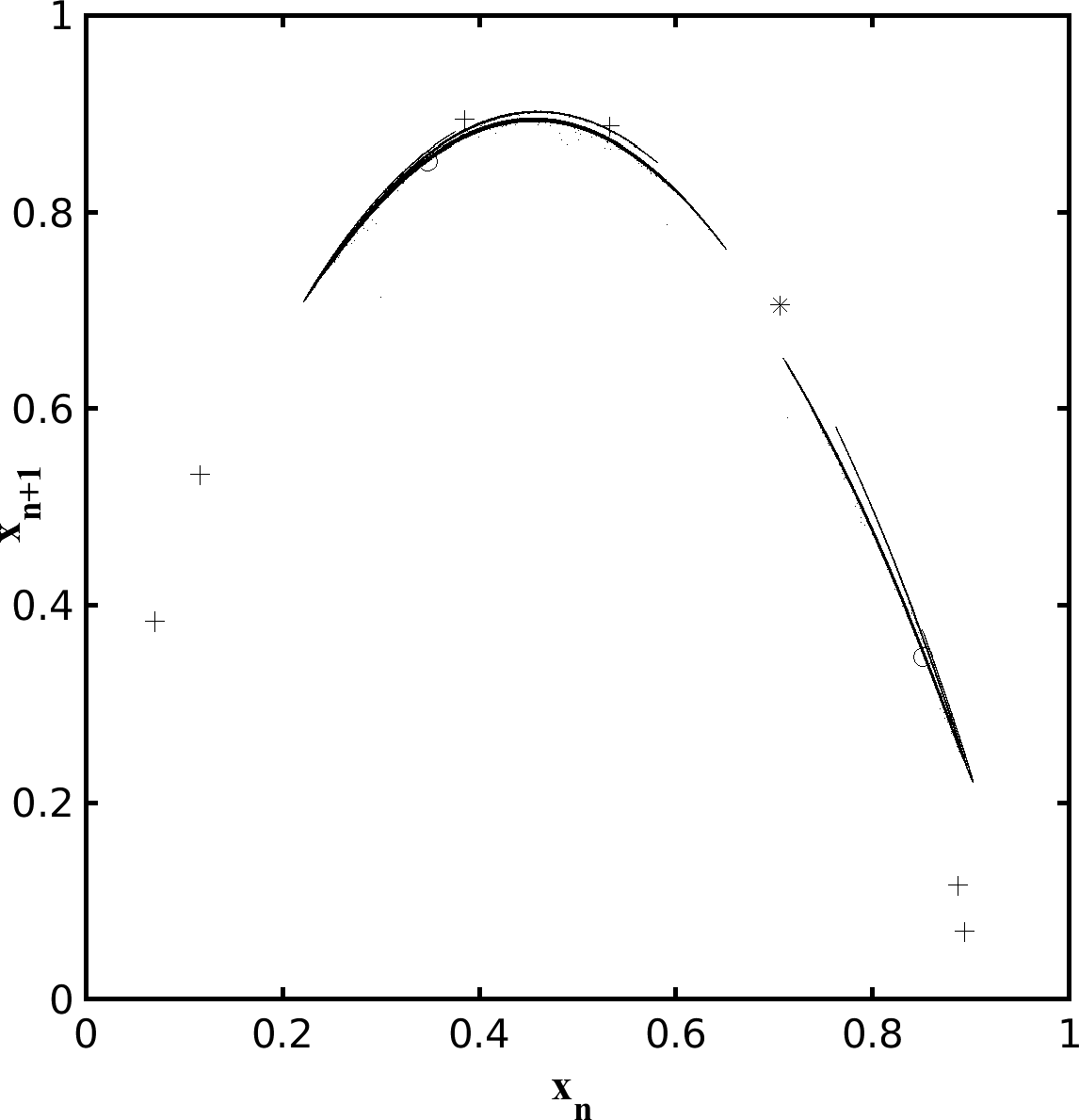}
\vspace{-0.25cm}
\caption{ The Poincar\'{e} plot (500000 iterations) for fractional difference logistic map with $\alpha=0.75$, $K=3.4$, $h=1$, and the initial point $x_0=0.3$. The asymptotically unstable $T=2$ sink, \{0.348,0.852\} is marked by the circles and the unstable fixed point $(K-1)/K=0.706$ by the star. Two asymptotically unstable $T=3$ cycles, \{0.116,0.533,0.887\} and \{0.0696,0.385,0.894\}, are marked by the plus signs. }
\label{fig4}
\end{figure}

\subsection{Poincar\'{e} plots}
\label{sec:7.2}


\begin{figure}[!t]
\includegraphics[width=0.45 \textwidth]{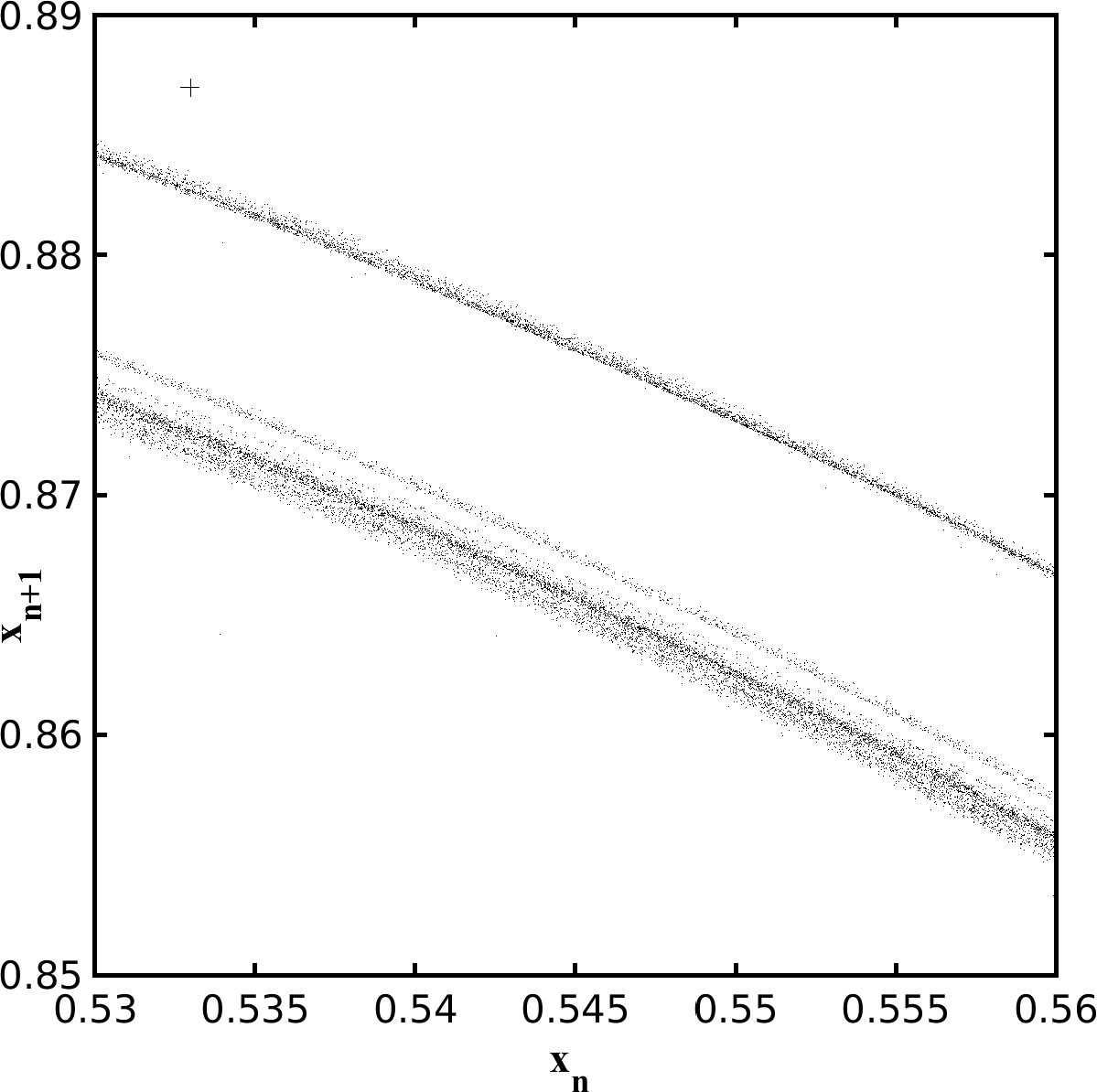}
\vspace{-0.25cm}
\caption{ Zoom of Fig.~\ref{fig4} with $\alpha=0.75$, $K=3.4$, $h=1$, and the initial point $x_0=0.3$. }
\label{fig5}
\end{figure}

Poincar\'{e} plots (return maps) present a good instrument to analyze cyclic behavior and chaos in fractional/fractional difference maps. Return maps for the fractional difference logistic map with $\alpha=0.75$, various values of $K$, and the initial point $x_0=0.3$, obtained after 500000 iterations, are given in Figs.~\ref{fig2}--\ref{fig4}. Calculated using Eq.~(\ref{eqT2logSolu}), $T=2$ sink is stable in Fig.~\ref{fig2} and unstable in Figs.~\ref{fig3}~and~\ref{fig4}. It is easy to see that the return map in Fig.~\ref{fig2} converges to the $T=2$ point calculated using Eqs.~(\ref{LimDifferences})~and~(\ref{Close}). This confirms the correctness of the equations and our calculations. 

\section{Chaos in discrete fractional systems}
\label{sec:8}

The chaotic return map in Fig.~\ref{fig4} looks like a multi-scroll attractor of a dissipative system. Similarity of the behavior of fractional systems to the behavior of dissipative systems was noticed by many authors and one of the first examples of this similarity can be found in \cite{ZSE}. Fig.~\ref{fig5} demonstrates a quasi-fractal structure of the Poincar\'{e} plot in Fig.~\ref{fig4}.
Chaos in fractional systems was investigated in many papers (see, e.g., \cite{ME3,ME2,Chaos,ME4,HBV4,Chaos2018,Chaos2018,ME9,ZSE,WBLya,RP}). In those papers one may find bifurcation diagrams and various forms of fractional chaotic attractors, including some new forms, for example, cascade of bifurcation type trajectories. Finding Lyapunov exponents \cite{WBLya} in the case of fractional maps is complicated and not always practical because converging trajectories may follow the power law and chaotic trajectories may first converge to a periodic trajectories which break after many iterations (see, e.g., \cite{ME11}). A consistent quantitative analysis of chaos in discrete fractional systems is still an open problem.

In regular maps the universal period doubling route to chaos is a well qualitatively and quantitatively established fact (see Mitchell Feigenbaum's paper \cite{Fei} and a selection of papers compiled by Predrag Cvitanovic \cite{CU}). Numerical results obtained in multiple papers show that the same qualitative property, the universal period doubling route to chaos (see, e.g. \cite{ME10}), is valid for fractional maps. Asymptotically periodic trajectories of all particular (depending on the function $G_K(x)$) implementations of the universal fractional maps, equations Eqs.~(\ref{FrCMapx})~and~(\ref{FalFacMap_h}), are defined by equations 
Eq.~(\ref{LimDifferences}) and~Eq.~(\ref{Close}).
Numerical simulations show that, like in nonlinear maps without memory,
for certain values of fractional map's parameter $K<K_{n}$ (or $K>K_{n}$), $n\in\mathbb{N}$, the map has a stable $T=2^{n-1}$-cycle.
This cycle is a real solution of the $2^n$ equations that define $T=2^{n}$-cycles and there are no real solutions for $T=2^{n}$-cycles which are not $T=2^{n-1}$-cycles. At the point $K=K_{n}$ a transition from complex to real solutions occurs and a stable real $T=2^{n}$-cycle, which is not a $T=2^{n-1}$-cycle, is born. At this point the $T=2^{n-1}$-cycle becomes unstable.
The nontrivial fact, which requires a thorough investigation, is the {\bf{conjecture}} that the $2^n$ equations Eq.~(\ref{LimDifferences})~and~Eq.~(\ref{Close}) that define asymptotically $2^{n}$-cycles have the same real solutions as the $2^{n-1}$ equations that define $2^{n-1}$-cycles for the values of $K$ up to $K=K_{n}$, at which point a transition from complex to real solutions occurs and these equations acquire a set of $2^n$ real solutions.   
Related to this conjecture is another {\bf{conjecture}} that for fractional maps there is a universal limit, which may depend on $\alpha$, similar to the Feigenbaum's constant
\begin{equation}
\delta=\lim_{n \rightarrow \infty}\frac{K_{n-1}-K_{n-2}}{K_{n}-K_{n-1}}.
\label{Univ}
\end{equation}

\section{Conclusion}
\label{sec:9}

The main idea of this paper is to use cyclic sinks to analyze periodic behavior and chaos in discrete fractional systems. The first step of this analysis is an algorithm to calculate the cyclic points. In this paper we derived equations, Eq.~(\ref{LimDifferences}) and  Eq.~(\ref{Close}), which define asymptotically cyclic points. These equations contain coefficients (sums $S_i$) which are the same for all maps. We calculated (and tabulated) these coefficients for period $T=2$ and $T=3$ cycles. Similar calculations can be used to calculate $S_i$ for any periodic cycles.

We also constructed Poincar\'{e} plots (return maps) for the logistic fractional difference family of maps. The plots confirm the correctness of the analysis done in this paper. We propose and plan to use periodic sinks to analyze the period doubling route to chaos and chaotic behavior of fractional systems. In the following papers we plan to add higher order unstable cyclic points to Poincar\'{e} plots of fractional chaotic systems in order to investigate regularities in chaotic attractor-like structures of Poincar\'{e} plots of fractional systems.

\begin{acknowledgements}
The author acknowledges continuing support from Yeshiva University 
and expresses his gratitude to the administration of Courant
Institute of Mathematical Sciences at NYU
for the opportunity to complete all computations at Courant. 
\end{acknowledgements}

\section*{Conflict of interest}
The authors declare that they have no conflict of interest.


\end{document}